# Generalization of electrical resistance scaling to Dirac fields on perforated fractals

Jonathan F. Schonfeld
Center for Astrophysics | Harvard and Smithsonian
60 Garden St., Cambridge, Massachusetts 02138 USA
jschonfeld@cfa.harvard.edu
ORCID ID# 0000-0002-8909-2401

**Abstract** I construct perforated, "take-away" fractals that support short-distance power-law scaling *with complex exponents* for Dirac (spin-1/2) propagators. The construction relies on a fortuitous ansatz for Dirac boundary conditions at the surfaces of spherical voids in a three-dimensional embedding space, and requires that boundary conditions vary from void to void and are distributed statistically. The appropriate distribution can ensure that nonzero charges (but not dipoles, which drive nontrivial power-law scaling) induced in voids cancel out in spatial averaging.

1. **Introduction**

Over the last four decades, a substantial literature has developed around the short-distance scaling behavior of point-to-point electrical resistance on fractals [1-3]. This is obviously important for the material properties of real media. But it also has broader mathematical significance, because point-to-point resistance is equivalent to the electrical potential of a point charge, and that in turn is equivalent to the Green's function of the time-independent free scalar wave equation.

Green's functions (also referred to as propagators) of all sorts of free wave equations are at the core of quantum field theory, which is the mathematical foundation for all models of elementary particles. The short-distance singular power-law behavior of these propagators is pivotal for renormalizability, i.e. the feasibility of extracting finite physical predictions for elementary particle processes. In the dimensional regularization technique [4], the short-distance scaling powers of free propagators are artificially softened so that otherwise divergent integrals in perturbative quantum amplitudes become finite. Over the years, a number of papers, including references [5-9], have proposed that the short-distance scaling of free propagators could be softened in a more fundamental and physically intuitive fashion if space itself were a fractal. Some of this work is axiomatic; some is constructive, focusing largely on lattice approximations to fractal spaces.

In Reference [10] I explicitly constructed a randomly perforated, or "take-away" fractal, on which the propagator of a free scalar field would exhibit softened short-distance power-law scaling. The "take-away" fractal was specifically defined by removing small voids from a three-dimensional "starting space." I did this for several reasons: First, because propagators of any spin can easily be defined in Euclidean 3-space, and a robust literature on wave-equation boundary conditions ([11-13] shows examples for Dirac fields and the general-relativity metric; references for scalar and Maxwell fields are too numerous to cite) provides tools for modifying the propagators around voids. And second, because distributing the voids randomly exempts us from the complications of accidental crystallographic symmetries in lattice models, and makes intuitive some important considerations that would otherwise be quite hard to think about.

But to realize the benefits of dimensional regularization through a physical model, it is not enough to soften short-distance power-law scaling, because the dimensional regularization technique also involves analytic continuation of scaling exponents *into the complex domain*. Without this continuation, dimensional regularization only makes so-called "logarithmic divergences" convergent, but those are not the only significant divergences in quantum field theory. To be sure, there is a literature on "complex dimensions" (e.g., [14-16]), but it is only tangentially related to the complex scaling exponents that I focus on here. In particular, if power-law scaling with radius $r$ is a functional form $r^u$, then the literature on complex dimensions is concerned with cases in which $u$ is real and depends in an oscillating fashion on $\ln r$. For example, if $u = u_0 + u_1 \cos(\omega \ln r)$, then for small $u_1$ we have

$$r^u \sim r^{u_0} + \frac{u_1}{2} r^{u_0 + i\omega} + \frac{u_1}{2} r^{u_0 - i\omega} \tag{1.1}$$

and the exponents in the two order-$u_1$ corrections are (up to normalization) referred to as complex dimensions. By contrast, in the present paper we are concerned with cases in which *the leading exponent $u_0$* itself is manifestly complex. There appears to be no literature on this, aside from Reference [17].

In Reference [17], I proposed several ways, based on random take-away-fractal constructions, to achieve complex scaling exponents for scalar-field propagators, all involving adjustment of boundary conditions at the surfaces of voids (and inclusion of complex terms in the quantum Lagrangian). I also attempted to do the same for Maxwell and Dirac fields. In the Dirac (spin-1/2) case, I was unable to identify boundary conditions that endow propagators with nontrivial power-law scaling when voids are impenetrable, but I was able to do so for "leaky" voids, albeit in a perturbative limit beyond which I didn't know how to go.

The present paper introduces boundary conditions that do in fact lead to complex power-law scaling in Dirac propagators on perforated fractals (with either impenetrable or leaky voids) *without* recourse to a perturbative limit. The fractals have spherical voids and are geometrically unexceptional, but boundary conditions at the surfaces of their voids vary from void to void and are distributed statistically (the idea of a statistical distribution of boundary conditions was first introduced in Reference [10], for scalar fields). This is necessary because Dirac boundary conditions induce not just dipoles (which drive nontrivial power-law scaling) in voids, but also nonzero total charges (which can make a Dirac field effectively massive). A statistical distribution of boundary conditions can guarantee that nonzero induced charges – but not dipoles – cancel out in spatial averaging. (On the other hand, maybe one might want a residual total charge – and therefore effective mass – to arise from the fractal geometry of space, but there is nothing in present-day particle physics which seems to cry out for it.[1]) A statistical distribution of boundary conditions can also produce scaling exponents that are complex. This is worth knowing about, because it is clearly advantageous for non-renormalizable field theories, although it by no means resolves all their problems.

---

[1]Statistical distribution of boundary conditions doesn't always help mitigate induced charges. In Ref [17] I sought to impose Robin boundary conditions on a scalar field at the surfaces of impenetrable voids by supplementing the bulk field Lagrangian with a boundary term $\frac{\beta}{2} \oiint \phi^2$. But that induced nonzero total charges in the voids, and it's not clear that averaging over various values of $\beta$ would help mitigate. I could have avoided inducing nonzero charge into impenetrable voids altogether by using the alternative surface term $\frac{\beta}{2} \oiint (\partial \phi / \partial n)^2$, where *n* is displacement normal to the surface. In that case, the boundary conditions become $\frac{\partial \phi}{\partial n} + \beta \phi =$ floating constant and $\oiint \frac{\partial \phi}{\partial n} = 0$, and induced charge is completely obviated.

This paper is organized as follows. In the next section, we review the basic facts about take-away fractals embedded in three-dimensional Euclidean space (extension to four-dimensional Minkowski is discussed in Reference [10] and will not be further considered here). This includes a summary of how dipole moments of voids lead to nontrivial short-distance power-law scaling. In Section 3, we present and build on an ansatz for Dirac-equation boundary conditions that produces simple and exact solutions for the field induced by an impenetrable sphere in a uniform background. Conclusions are summarize in Section 4.

## 2. Take-Away Fractals

The content of this section paraphrases material from References [10] and [18].

A random "take-away" fractal is a set formed by the following recursive procedure: For definiteness, start with three-dimensional Euclidean space, and a reference block of volume $V$. Distribute points throughout space randomly with density $\rho$, and, centered at each such point, remove a copy of the reference block. Call this process the zeroth iteration and its result the zeroth prefractal. Now choose an arbitrary scale factor $\xi>1$ and define the $k$'th iteration inductively as follows:

- Distribute points randomly with density $\rho\xi^{3k}$ throughout whatever part of the Euclidean space has not been removed by preceding iterations.
- Centered at each such point, remove from the $k$-1'st iteration a copy of the reference block linearly scaled by factor $\xi^{-k}$.

The result is the $k$'th prefractal. In the limit of infinite $k$, what's left has fractal dimension $D=3+\ln(1-\rho V)/\ln\xi$. We refer to the term $\rho V$ as void footprint; the factor $(1-\rho V)$ is the volumetric proportion of iteration $k$-1 removed by iteration $k$; and the ratio of logarithms is minus a dimension deficit for physical $\rho V<1$. Figure 1 illustrates the first three iterations of such a construction starting in two-dimensional Euclidean space with circular voids having $\rho V \sim 1/9$ and $\xi \sim 3$.

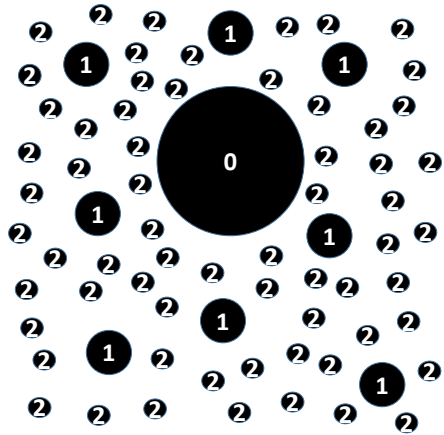

Figure 1: Cartoon of first three iterations of random take-away fractal construction starting with two-dimensional Euclidean space, for circular voids having $\rho V \sim 1/9$ and $\xi \sim 3$. Voids are labeled by the iteration numbers at which they're removed. The resulting prefractal for $k$=2 is the space between voids.

James Clerk Maxwell was the first to recognize [19] that the Green's function of a point source in a take-away prefractal brings to mind the potential of a point charge in a dielectric, if we can analogize atoms in the dielectric to voids in the prefractal. Voids distort an ambient potential, and far from the void, the distortion is dominated by an induced dipole moment. So, for sparse enough voids (small $\rho V$), the analogy to a dielectric composed of polarizable atoms should be good. According to dielectric theory (Reference [20], Eq. (4.74) et seq.), voids removed in iteration $k$ shield or amplify an ambient field by a factor

$$\Phi_k = \left[1 + \frac{4\pi\rho\xi^{3k}p_k}{1 - \frac{4\pi}{3}\rho\xi^{3k}p_k}\right]^{-1}, \tag{2.1}$$

where $p_k$ is a void's polarizability. It's convenient to write $p_k = (3V/4\pi\xi^{3k})g$; then for spherical voids, $g=+1$ for Dirichlet boundary conditions and $-1/2$ for Neumann [10], either of which makes Expression (2.1) independent of $k$; accordingly, we refer simply to $\Phi$ without a subscript.

These factors compound for different values of $k$. Each take-away iteration multiplies the force (gradient of potential) of a point charge by $\Phi$ in the space between voids, but only for iterations whose voids are *smaller* than the distance to the point charge, since larger voids don't fit. Thus, we can write

$$Force \propto r^{-2}\Phi^{-k_{max}}, \tag{2.2}$$

up to normalization, where $r$ is distance from the point charge, and $k_{max}$ is the highest iteration whose removed spheres are larger than or equal to $r$ in linear dimension. Since $k_{max}$ satisfies $r \sim$ radius of iteration-$k_{max}$ sphere, equal to $R/\xi^{k_{max}}$ ($R$=radius of reference spherical void), expression (2.2) amounts to power-law scaling of the form

$$r^{-1+\ln\Phi/\ln\xi} \tag{2.3}$$

for the potential.

## 3. Solvable Dirac Boundary Conditions at Impenetrable Surface

For electrostatics, one uses the familiar method of images [20] to construct the potential induced outside a sphere due to a uniform background (uniform potential or gradient). This amounts to an ansatz that there is an image dipole and possibly an image point charge at the center of the sphere. Our goal in this section is to extend this method of image charges to a Dirac field in a uniform background.

The Dirac equation [21] restricted to Euclidean 3-space (i.e. time-independent Dirac equation) is

$$\boldsymbol{\gamma} \cdot \boldsymbol{\nabla}\psi = \eta, \tag{3.1}$$

where $\psi$ is a four-component complex spinor field, $\eta$ is a four-component complex spinor density, and $\boldsymbol{\gamma}$ is a three-vector of 4x4 anti-Hermitian matrices that satisfy

$$\{\gamma_i, \gamma_j\} = -\delta_{ij} \tag{3.2}$$

such that there is a fourth Hermitian matrix $\gamma_0$ that anticommutes with all the $\gamma_i$ and satisfies $\gamma_0^2=1$. A Dirac point charge at the origin corresponds to the field

$$\psi = \frac{\boldsymbol{\gamma} \cdot \hat{\boldsymbol{r}}}{r^2} \chi \tag{3.3}$$

where $\chi$ is a fixed complex spinor. A Dirac dipole, derived by differentiating Expression (3.3) with respect to position, corresponds to the field

$$\psi = \boldsymbol{\gamma} \cdot \frac{\boldsymbol{d} - 3\hat{\boldsymbol{r}}(\boldsymbol{d} \cdot \hat{\boldsymbol{r}})}{r^3} \tag{3.4}$$

where $\boldsymbol{d}$ is a three-vector of fixed complex spinors. One determines $\chi$ and $\boldsymbol{d}$ by adding the background field to the forms (3.3) and (3.4) and then fitting the result to boundary conditions at the surface of the sphere.

The boundary condition for a time-independent Dirac field at an impenetrable static surface ([11, 12]) is specified by a unitary Hermitian 4x4 matrix $M$ that satisfies

$$\{M, \gamma_0 \boldsymbol{n} \cdot \boldsymbol{\gamma}\} = 0, \tag{3.5}$$

where $\boldsymbol{n}$ is the unit normal to the surface, and the matrix $M$ can vary over the surface. The boundary condition itself is

$$(1 - M)\psi = 0 \tag{3.6}$$

or equivalently

$$\psi = (1 + M)\psi_0 \tag{3.7}$$

for some arbitrary $\psi_0$. (We address penetrable surfaces later.)

Any such matrix $M$ satisfies [17]

$$M = a_1 \gamma_0 + a_2 i \boldsymbol{n} \cdot \boldsymbol{\gamma} + a_3 \gamma_0 \boldsymbol{n}_\perp \cdot \boldsymbol{\gamma} + a_4 \gamma_0 \boldsymbol{n}_{\perp\prime} \cdot \boldsymbol{\gamma} + a_5 i \gamma_5 \gamma_0 + a_6 \gamma_5 \boldsymbol{n} \cdot \boldsymbol{\gamma} + a_7 \gamma_5 \gamma_0 \boldsymbol{n}_\perp \cdot \boldsymbol{\gamma} \\ + a_8 \gamma_5 \gamma_0 \boldsymbol{n}_{\perp\prime} \cdot \boldsymbol{\gamma}, \tag{3.8}$$

for some real numbers $a_i$, where the orthogonal unit vectors $\mathbf{n}_\perp$ and $\mathbf{n}_{\perp'}$ are defined in terms of an arbitrary reference unit vector $\mathbf{q}$ by

$$\mathbf{n}_\perp \equiv (\mathbf{q} - \mathbf{n}(\mathbf{q} \cdot \mathbf{n}))/\sqrt{1 - (\mathbf{q} \cdot \mathbf{n})^2}, \quad \mathbf{n}_{\perp'} \equiv (\mathbf{q} \times \mathbf{n})/\sqrt{1 - (\mathbf{q} \cdot \mathbf{n})^2}, \tag{3.9}$$

and where $\gamma_5 \equiv i\gamma_0\gamma_1\gamma_2\gamma_3$. Two easy classes of solutions for $M$ are

$$M = a_1\gamma_0 + a_3\gamma_0\mathbf{n}_\perp \cdot \boldsymbol{\gamma} + a_4\gamma_0\mathbf{n}_{\perp'} \cdot \boldsymbol{\gamma} + a_5 i\gamma_5\gamma_0, \tag{3.10}$$

$$M = a_2 i\mathbf{n} \cdot \boldsymbol{\gamma} + a_6\gamma_5\mathbf{n} \cdot \boldsymbol{\gamma} + a_7\gamma_5\gamma_0\mathbf{n}_\perp \cdot \boldsymbol{\gamma} + a_8\gamma_5\gamma_0\mathbf{n}_{\perp'} \cdot \boldsymbol{\gamma}, \tag{3.11}$$

where either expression is normalized by having all its $a_i^2$ sum to unity. (One can also make a hybrid by taking either form of $M$ and transforming it according to

$$[\exp(-i\phi\gamma_0\mathbf{n} \cdot \boldsymbol{\gamma})]M[\exp(i\phi\gamma_0\mathbf{n} \cdot \boldsymbol{\gamma})] \tag{3.12}$$

for real $\phi$.)

The square root in the denominators of equation (3.9), together with the normalization conditions for Equation (3.10) or (3.11), would seem a practical obstacle to fitting point charge (3.3) and point dipole (3.4) to these boundary conditions. But we can eliminate this obstacle with a simple ansatz: In Equation (3.11), set

$$a_8 = 0, a_7 = as\sqrt{1 - (\mathbf{q} \cdot \mathbf{n})^2}, a_6 = a(\mathbf{q} \cdot \mathbf{n}), a_2 = b \tag{3.13}$$

for some real constant $a$ between -1 and +1, real $b$ with $a^2+b^2=1$, and $s=\pm 1$ (we will address the anomalous role of $a_8$ momentarily). If we insert this into Equation (3.7) with constant $\psi_0$, then, on the surface of an impenetrable sphere, we have

$$\psi = \left[1 + bi\hat{\mathbf{r}} \cdot \boldsymbol{\gamma} + a(\mathbf{q} \cdot \hat{\mathbf{r}})\gamma_5\hat{\mathbf{r}} \cdot \boldsymbol{\gamma} + as\gamma_5\gamma_0(\mathbf{q} - \hat{\mathbf{r}}(\mathbf{q} \cdot \hat{\mathbf{r}})) \cdot \boldsymbol{\gamma}\right]\psi_0. \tag{3.14}$$

It is easy to rearrange Equation (3.14) so that it becomes equivalent to the field on the surface of a sphere of radius $Q$ due to the sum of background field, dipole and point charge:

$$\psi_B = \left[1 + \frac{1}{3}a\gamma_5(1 - 2s\gamma_0)\mathbf{q} \cdot \boldsymbol{\gamma}\right]\psi_0 \tag{3.15}$$

$$\mathbf{d} = \frac{1}{3}aQ^3\gamma_5(s\gamma_0 + 1)\mathbf{q}\psi_0 \equiv Q^3\mathbf{U}\mathbf{q}\psi_0 \tag{3.16}$$

$$\chi = iQ^2 b\psi_0. \tag{3.17}$$

Since $\psi_B$ is the imposed field, this isn't complete without inverting Equation (3.15):

$$\psi_0 = \left[1 - \frac{1}{3}a\gamma_5(1-2s\gamma_0)\mathbf{q}\cdot\boldsymbol{\gamma}\right]\left\{\frac{9-a^2(2-s\gamma_0)^2}{(9-a^2)(1-a^2)}\right\}\psi_B \equiv T\psi_B. \tag{3.18}$$

(We see why it was necessary to introduce nonzero $b$: because otherwise $a=1$ and Equation (3.18) is singular.) We see directly that if voids all have the same value of $a$, but come in two equal-population varieties with $b=\pm(1-a^2)^{1/2}$, then the only collective effect that survives is due to induced dipoles and not point charges. The picture is similar if we generalize ansatz (3.14) to

$$\psi = \left[1 + bie^{i\gamma_5\theta}\hat{\mathbf{r}}\cdot\boldsymbol{\gamma} + a(\mathbf{q}\cdot\hat{\mathbf{r}})\gamma_5 e^{i\gamma_5\theta}\hat{\mathbf{r}}\cdot\boldsymbol{\gamma} + as\gamma_5\gamma_0(\mathbf{q} - \hat{\mathbf{r}}(\mathbf{q}\cdot\hat{\mathbf{r}}))\cdot\boldsymbol{\gamma}\right]\psi_0 \tag{3.19}$$

for arbitrary real $\theta$.

In ansatz (3.13) we chose to avoid nonzero $a_8$ because that would have led to a term

$$\gamma_5\gamma_0(\mathbf{q}\times\mathbf{n})\cdot\boldsymbol{\gamma}\psi_0 = \mathbf{n}\cdot\mathbf{D}\psi_0 = \left[\mathbf{n}\cdot\mathbf{D} + \frac{1}{3}(\mathbf{n}\cdot\boldsymbol{\gamma})(\boldsymbol{\gamma}\cdot\mathbf{D})\right]\psi_0 - \frac{1}{3}(\mathbf{n}\cdot\boldsymbol{\gamma})(\boldsymbol{\gamma}\cdot\mathbf{D})\psi_0 \tag{3.20}$$

in the surface field, where

$$\mathbf{D} \equiv \gamma_5\gamma_0(\boldsymbol{\gamma}\times\mathbf{q}). \tag{3.21}$$

The last term in the far right-hand side of Equation (3.20) continues off the sphere as a Dirac point source at the origin. But, by construction, the term with square brackets continues off the sphere as a solution of the source-free Dirac equation (3.1) that grows linearly throughout space. We could pursue boundary solutions that match not only a background field but also its gradient, but I don't see how that adds distinct insight to the present paper.

With Equation (3.10) we could have started with the ansatz

$$a_4 = 0, a_3 = as\sqrt{1-(\mathbf{q}\cdot\mathbf{n})^2}, a_1 = a(\mathbf{q}\cdot\mathbf{n}), a_5 = b, \tag{3.22}$$

or

$$\psi = \left[1 + a(\mathbf{q}\cdot\mathbf{n})\gamma_0 + as\gamma_0(\mathbf{q}-\mathbf{n}(\mathbf{q}\cdot\mathbf{n}))\cdot\boldsymbol{\gamma} + bi\gamma_5\gamma_0\right]\psi_0. \tag{3.23}$$

But this once again leads to linear behavior at large distances, by analogy with Equation (3.20), substituting $\gamma_0\mathbf{q}$ for $\mathbf{D}$.

We now look specifically at the power law scaling implied by Eqs. (3.4), (3.16) and (3.18). First note that **d** is clearly equal to **q** times a single 4-matrix ($Q^3UT$) applied to $\psi_B$. It follows that the sum of all dipole contributions (3.4) is $Q^3UT\psi_B$ multiplied on its left by $\gamma$ dotted into the sum of

$$\frac{\mathbf{q} - 3\hat{\mathbf{r}}(\mathbf{q} \cdot \hat{\mathbf{r}})}{r^3} \tag{3.24}$$

over all voids at iteration $k$. By symmetry, this must be proportional to **q**. It follows trivially that the collective effect of dipoles (3.16) on a background Dirac field $\psi_B$ is the same as the collective effect of conventional electric dipoles on a background electrostatic field, i.e. matrix multiplication by the form (2.1) with

$$g = (\boldsymbol{\gamma} \cdot \mathbf{q})UT. \tag{3.25}$$

So, for small $\rho V$, we have

$$\ln \Phi \sim -3\rho V (\boldsymbol{\gamma} \cdot \mathbf{q})UT. \tag{3.26}$$

Clearly then we have four scaling laws (2.3), depending on the four eigenvalues of $-3\rho V(\boldsymbol{\gamma}\cdot\mathbf{q})UT$. If all the eigenvalues were real, renormalizability would require that they all must be positive, in order to soften the canonical $1/r$ small-distance behavior. It is easy to simplify Equation (3.26) to

$$-3\rho V(\boldsymbol{\gamma} \cdot \mathbf{q})UT = \left(\frac{\rho V a}{1 - a^2}\right)[(\boldsymbol{\gamma} \cdot \mathbf{q})\gamma_5 - a](1 + s\gamma_0). \tag{3.27}$$

Using specific properties of gamma matrices [22], it's easy to show the four eigenvalues are

$$\left(\frac{\rho V a}{1 - a^2}\right)\left[-a \pm \sqrt{a^2 s^2 - (1 - s^2)}\right], \tag{3.28}$$

where each $\pm$ option is counted twice, and where we have withheld setting $s^2=1$ to leave open the possibility of averaging over a statistical ensemble of $s$ values. For a constant-$s$ (with $|s|=1$) "monoculture," clearly the eigenvalues are zero or $-2\rho V a^2/(1-a^2)$ and therefore never positive. For voids split equally between $s=\pm 1$, we should use the averaged value $s=0$ in Equation (3.28), resulting in eigenvalues that are all complex,

$$\left(\frac{\rho V a}{1 - a^2}\right)[-a \pm i], \tag{3.29}$$

which gives rise to the complex scaling that dimensional regularization tells us is desirable, although different spinor components have different complex scaling exponents.

The thought process in this Section can also be applied in a straightforward fashion to the leaky boundary conditions in Reference [17], with similar results.

## 4. Conclusion

We have presented a fractal construction that produces complex short-distance scaling exponents for free Dirac-field propagators in spaces of dimensional slightly less than three. This may be advantageous for making sense of some non-renormalizable quantum field theories, although it by no means resolves all their problems. This work is distinguished from most other papers about quantum fields on fractals in that it focuses on random take-away fractals. It is distinguished from the literature on "complex dimensions" in that it focuses on scaling exponents that are complex *even disregarding* terms that oscillate.

**Acknowledgments**

I am grateful to the referees for helpful feedback.

**Funding**

The author declares that no funds, grants, or other support were received during the preparation of this manuscript.

**Competing interests**

The author has no relevant financial or non-financial interests to disclose.